\documentclass[aps,nofootinbib,superscriptaddress,showpacs]{revtex4}
\usepackage{epsfig}
\usepackage[latin1]{inputenc}
\usepackage{float,amsmath,amstext,amsthm,amssymb,amsfonts,slashed}
\usepackage{graphicx}

\begin{document}

\title{A model of the effect of collisions 
       on QCD plasma instabilities}

\author{Bj\"orn Schenke}
\affiliation{Institut f\"ur Theoretische Physik \\
  Johann Wolfgang Goethe - Universit\"at Frankfurt \\ 
  Max-von-Laue-Stra\ss{}e~1, 
  D-60438 Frankfurt am Main, Germany\vspace*{5mm}}
\author{Michael Strickland}
\affiliation{Frankfurt Institute for Advanced Studies \\
  Johann Wolfgang Goethe - Universit\"at Frankfurt \\
  Max-von-Laue-Stra\ss{}e~1,
  D-60438 Frankfurt am Main, Germany\vspace*{5mm}}
\author{Carsten Greiner}
\affiliation{Institut f\"ur Theoretische Physik \\
  Johann Wolfgang Goethe - Universit\"at Frankfurt \\ 
  Max-von-Laue-Stra\ss{}e~1, 
  D-60438 Frankfurt am Main, Germany\vspace*{5mm}}
\author{Markus H. Thoma}
\affiliation{Max-Planck-Institute for Extraterrestrial
  Physics \\
  P.O. Box 1312, 85741 Garching, Germany}

\begin{abstract}
We study the effect of including a BGK collisional kernel on the 
collective modes of a QCD plasma which has a hard-particle distribution function 
which is anisotropic in momentum space.  We calculate dispersion 
relations for both the stable and unstable modes and show that the 
addition of hard particle collisions slows the rate of growth of QCD 
plasma unstable modes.  We also show that for any anisotropy there is 
an upper limit on the collisional frequency beyond which no 
instabilities exist.  Estimating a realistic value for the collisional 
frequency for $\alpha_s \sim 0.2 - 0.4$ we find that for the large-anisotropy 
case which is relevant for the initial state of matter generated by 
free streaming in heavy-ion collisions that the collisional frequency 
is below this critical value.
\end{abstract}



\maketitle


\section{Introduction and Motivation}
\label{Intro}

The ultrarelativistic heavy ion collision experiments ongoing at the Relativistic
Heavy Ion Collider (RHIC) at Brookhaven National Laboratory (BNL)
and planned at the Large Hadron Collider (LHC) at CERN study nuclear matter
under extreme conditions. A main point of interest is the
identification and investigation of a phase transition to the
quark-gluon plasma (QGP). If it is created, the QGP is expected to
expand, cool and finally hadronize. Whether or not it is possible
to apply a thermodynamic description of the created system in its
later stages of evolution depends on whether it thermalizes fast
enough. RHIC data indicate that such a rapid thermalization does
occur \cite{Heinz:2002un,Heinz:2002gs,Gyulassy:2004zy}, in contradiction
to leading order perturbation theory estimates. Therefore, many
recent works have been dedicated to the explanation of how the fast 
thermalization is achieved. One possibility is the assumption of a strongly
coupled QGP \cite{Thoma:2004sp,Thoma:2005uv}. In other approaches \cite{Kharzeev:2005iz} the process
of particle production leads to momentum distributions of the equilibrium form
immediately, without any secondary processes needed. However it is not explained how
the equilibrium state is maintained when the system is driven out of equilibrium by free streaming.
Secondary processes are certainly necessary to explain this. 
Within recent transport theory approaches, the inclusion of particle production and absorption via 
$2\leftrightarrow3$ pQCD bremsstrahlung processes speeds up the equilibration significantly \cite{Xu:2004mz} 
as compared to equilibration solely driven by perturbative binary collisions. 

Here we will concentrate on the role of non-equilibrium QCD collective modes in the
isotropization and thermalization of the system as investigated in
\cite{Mrowczynski:1993qm,Mrowczynski:1994xv,Mrowczynski:1996vh,Mrowczynski:2000ed,Birse:2003qp,Randrup:2003cw,Romatschke:2003ms,Arnold:2003rq,Romatschke:2004jh,Mrowczynski:2004kv,Dumitru:2005gp,Manuel:2005mx,Arnold:2005vb,Rebhan:2005re,Romatschke:2005pm} 
(for a recent review see \cite{Mrowczynski:2005ki}). 
In these approaches the QGP is assumed to be homogeneous and stationary, but
anisotropic in momentum space, such that kinetic instabilities can occur.
They are initiated either by charge or current fluctuations. In the first case, 
the electric field is parallel to the wave vector $\mathbf{k}$ 
($\mathbf{E}\parallel\mathbf{k}$), while in the second case the field is perpendicular
to $\mathbf{k}$ ($\mathbf{E}\perp\mathbf{k}$). This is why the corresponding 
instabilities are called longitudinal and transverse, respectively.
Since the electric field plays a crucial role in the generation of longitudinal modes,
they are also called electric, while the transverse modes are
called magnetic. The magnetic mode known as the filamentation- or Weibel-instability \cite{Weibel:1959}
appears to be relevant for the QGP \cite{Mrowczynski:1993qm,Mrowczynski:1994xv,Mrowczynski:1996vh}, created in heavy ion collisions.

The instabilities are found to have a significant effect on the 
system's evolution, leading to a faster isotropization and 
equilibration. The equilibration due to instabilities only happens 
indirectly, because the instabilities driven isotropization is a mean-field 
reversible process, which does not produce entropy. However, 
parton momentum distributions are influenced by the isotropization, 
which speeds up the equilibration. Collisions, being responsible for 
the dissipation are needed to reach the equilibrium state of maximum 
entropy.  All calculations so far have been carried out at leading 
order in perturbation theory, such that collisions among the hard 
particles that enter at higher orders in $g$ could be neglected. 
However, in heavy ion collision experiments at the RHIC and LHC the 
couplings expected are of the order $\alpha_s \sim 0.2-0.4$ and higher 
order terms will be important.  Hence, collisions can not simply be 
neglected and their effect on the system's evolution, particularly on 
the collective modes, has to be investigated.  We expect that the inclusion
of collisional damping will decrease the instability growth rate, however,
one would like to make this statement quantitative and say precisely by
how much the growth rate is reduced.
This is the main objective of the present work in that we give a first quantitative 
estimate of how the collisions affect the dispersion relations of the 
collective modes and the growth of instabilities in particular. To 
achieve this, we introduce a model for the inclusion of collisions 
based on the Vlasov equations for QCD combined with a BGK-type \cite{Bhatnagar:1954}
collision term. The regarded anisotropic distributions are derived 
from an isotropic one by contracting the hard particle distribution 
function along a preferred direction. We concentrate on the case in 
which the direction of anisotropy is parallel to the wave vector of 
the regarded collective mode, because in this case the growth rate of 
the magnetic instability is maximal and analytic expressions for the 
structure functions can be found. For any such distribution we find 
that the growth rate of instabilities decreases approximately linearly 
with increasing collision rate and there exists a critical collision 
rate above which instabilities cease to exist.

The current work is organized as follows: In Section \ref{review}
we briefly review the kinetic approach in the collisionless case.
We introduce the model for the inclusion of collisions in Section
\ref{includecollisions}, which closes with the final result for
the self energy of the collective modes. Section \ref{dispersion}
briefly reviews the procedure for finding the dispersion
relations and in Section \ref{anisosystem} we derive the structure
functions and dispersion relations for the specified anisotropic
systems. Final results for the dispersion relations of the stable
modes including collisions are given in Section \ref{stable} while
the influence of collisions on the unstable modes is shown in 
Section \ref{unstable}.  In Section \ref{discussions} we
discuss further the results and attempt to estimate the numerical value of
the collisional frequency.  In Section \ref{conclusions} we conclude
and give an outlook.

\section{Collision free transport equations}
\label{review} In the kinetic theory approach
\cite{Elze:1989un,Mrowczynski:1993qm,Blaizot:1993zk,Mrowczynski:1994xv,Kelly:1994dh,Mrowczynski:1996vh,Blaizot:1999xk,Mrowczynski:2000ed,Blaizot:2001nr,Randrup:2003cw}
gluons, quarks and antiquarks are described by their phase space
densities, given by the gauge covariant Wigner functions
$n^{i}(p,X)$, with $i\in\{g,q,\bar{q}\}$.
To obtain the linearized transport equations we expand 
$n^{i}(p,X)=n^{i}(\mathbf{p})+\delta n^{i}(p,X)$. 
After a gauge covariant
gradient expansion of the generalized QCD-Kadanoff-Baym equations
for the quark and gluon Green functions, the linearized equations of motion read 
\cite{Elze:1989un,Mrowczynski:1993qm,Blaizot:1993zk,Mrowczynski:1994xv,Kelly:1994dh,Mrowczynski:1996vh,Blaizot:1999xk,Mrowczynski:2000ed,Blaizot:2001nr,Randrup:2003cw}
\begin{align}
[V\cdot D_X,\delta n^{i}(p,X)]& + g \theta_{i}
V_{\mu}F^{\mu\nu}(X)\partial_{\nu}^{(p)}n^{i}(\mathbf{p})=0\label{trans1}\text{\,,}
\end{align}
where $V=(1,\mathbf{v})$
with $\mathbf{v}=\mathbf{p}/|\mathbf{p}|$ and
$\theta_{g}=\theta_{q}=1$ and $\theta_{\bar{q}}=-1$. 
$D_X=\partial_X+igA(X)$ is
the covariant derivative with the soft gauge field
$A^{\mu}=A^{\mu}_a T^a$ or $A^{\mu}=A^{\mu}_a t^a$ in the gluon
and quark equations of motion, respectively. The gluon field
strength tensor is given by
$F^{\mu\nu}=\partial^{\mu}A^{\nu}-\partial^{\nu}A^{\mu}-ig[A^{\mu},A^{\nu}]$.
Eq. (\ref{trans1}) holds 
for each particle species, with the color neutral background fields 
$n^{q/\bar{q}}(\mathbf{p})=f^{q/\bar{q}}(\mathbf{p})I$ and
$n^{g}(\mathbf{p})=f^{g}(\mathbf{p})\mathcal{I}$. 
$I$ and $\mathcal{I}$ are unit matrices in the fundamental and adjoint representation, respectively.
The scalar functions $f^{q/\bar{q}}(\mathbf{p})$ and $f^{g}(\mathbf{p})$ are found by projection:
\begin{align}
	f^{q/\bar{q}}(\mathbf{p}) &= \frac{1}{N_c} \mathrm{Tr}\left[n^{q/\bar{q}}(p,X)\right] \text{\,,} \notag \\
	f^{g}(\mathbf{p}) &= \frac{1}{N_c^2-1} \mathrm{Tr}\left[n^{g}(p,X)\right]\text{\,.}
\end{align}
The induced fluctuations 
$\delta n^{q/\bar{q}}(p,X)=\delta f^{q/\bar{q}}_b(p,X)t^b$ and $\delta n^{g}(p,X)=\delta f^{g}_b(p,X)T^b$, with the generators
$t_b$ and $T_b$ in the fundamental and adjoint representation, respectively,
are assumed to be much smaller than the colorless background terms.
The scalar functions $\delta f^{i}_a(p,X)$ again follow by projection:
\begin{align}
	\delta f^{q/\bar{q}}_a(p,X) &= 2 \mathrm{Tr}\left[t_a n^{q/\bar{q}}(p,X)\right]\text{\,,} \notag\\
	\delta f^{g}_a(p,X) &= \frac{1}{N_c} \mathrm{Tr}\left[T_a n^{g}(p,X)\right]\text{\,.}
\end{align}

The induced current for each color channel reads \cite{Mrowczynski:2000ed,Blaizot:2001nr}
\begin{equation}
	J_{\text{ind}\,a}^{i\,\mu}=g \int_{\mathbf{p}} V^{\mu}\left\{2N_c \delta f^{g}_a(p,X)+N_{f}[\delta f^{q}_a(p,X)
	-\delta f^{\bar{q}}_a(p,X)]\right\}\,\text{,}\label{indcurrent}
\end{equation}
with 
\begin{equation}
	\int_{\mathbf{p}} := \int \frac{d^3p}{(2\pi)^3}\text{\,.}
\end{equation}
When we concentrate on the soft scale $k\sim gT\ll T$,
the first scale at which collective motion appears,
we can neglect terms of subleading order in $g$ and the theory becomes
effectively Abelian.\footnotemark \footnotetext{This strictly only occurs at leading order in $g$. The induced
current also contains terms of higher order in $g$ (and $A$), which correspond to the non-Abelian self-interactions
of the soft gauge field.  Here we are looking at the self-energy, so that we can ignore these.
}
Now all color channels decouple in Eq.~(\ref{trans1})
and we find separate solutions for each $\delta f^{i}_a(p,X)$.
After Fourier transformation the total
induced current (\ref{indcurrent}) becomes \cite{Mrowczynski:2000ed}
\begin{equation}
    J^{\mu}_{\text{ind}\,a}(K)=g^2 \int_{\mathbf{p}} V^{\mu}
    \partial^{\beta}_{(p)}
    f(\mathbf{p})\left(g_{\gamma\beta}-\frac{V_{\gamma}K_{\beta}}{K\cdot
    V+i\epsilon}\right)A^{\gamma}_a(K)+\mathcal{O}(g^3 A^2)
    \,\text{,}
\end{equation}
where $K=(\omega,\mathbf{k})$ and the effective background phase space density is given by
\begin{equation}
f(\mathbf{p})=
2N_c f^{g}(\mathbf{p})+N_f\left[f^{q}(\mathbf{p})+f^{\bar{q}}(\mathbf{p})\right] \label{f}\text{\,.}
\end{equation}
Functional differentiation of the induced current with respect to the gauge field $A_{\nu}$ 
and taking $A_{\nu}\rightarrow 0$ leads to the gluon self-energy,
which equals the one which can be obtained by a hard-loop resummation within the diagrammatic 
approach in the collisionless case. 


\section{Inclusion of collisions}
\label{includecollisions} 
In the effectively Abelian limit, introduced in Section \ref{review}, the
equation of motion (\ref{trans1}) holds for each color channel separately, such that we
can write
\begin{equation}
V\cdot \partial_X \delta f^{i}_a(p,X) + g \theta_{i}
V_{\mu}F^{\mu\nu}_a(X)\partial_{\nu}^{(p)}f^{i}(\mathbf{p})=\mathcal{C}^{i}_a(p,X)\label{trans2}\,\text{,}
\end{equation}
where we have included the collision term $\mathcal{C}^{i}_a$, given by
\begin{equation}
    \mathcal{C}^{i}_a(p,X)=-\nu\left[f^{i}_a(p,X)-\frac{N^{i}_a(X)}{N^{i}_{\text{eq}}}f^{i}_{\text{eq}}(|\mathbf{p}|)\right]\,\text{,}\label{collision}
\end{equation}
with $f^{i}_a(p,X)=f^{i}(\mathbf{p}) + \delta f^{i}_a(p,X)$. 
This BGK-type collision term \cite{Bhatnagar:1954} 
describes how collisions equilibrate the system within a time proportional to 
$\nu^{-1}$. 
Here we will assume that
the collision rate $\nu$ is independent of momentum and particle species, however,
these assumptions are easily relaxed.
Note that these collisions are not color-rotating.
The particle numbers are given by
\begin{align}
    N^{i}_a(X)=\int_{\mathbf{p}} f^{i}_a(p,X) \text{ , ~} N^{i}_{\text{eq}}=\int_{\mathbf{p}} f^{i}_{\text{eq}}(|\mathbf{p}|)=\int_{\mathbf{p}} f^{i}(\mathbf{p})\text{\,.}\label{numbers}
\end{align}
We note that the difference between the BGK collisional kernel (\ref{collision}) and
the conventional relaxation-time approximation (RTA) is the multiplication of the second term
in brackets by the ratio of the density over the equilibrium density.  RTA simply takes the 
difference of the distribution function and the equilibrium distribution function implicitly 
setting this ratio to one.
The advantage of the BGK kernel over an RTA kernel lies in the fact that the number
of particles is instantaneously conserved by the BGK collisional-kernel, 
i.e., that
\begin{equation}
	\int_{\mathbf{p}} \mathcal{C}^{i}_a(p,X)=0\text{\,.}
\end{equation}
This simply states that the collisions can only occur if a particle is present and that only the 
momentum of the particles will change as a result, not the particle number.  This condition is 
violated by RTA. In addition, Ref.~\cite{Manuel:2004gk} showed that after
a simple modification the BGK collisional-kernel covariantly conserves the
color current.

The inclusion of a BGK collisional kernel allows for simulation of the 
effect of binary collisions with substantial momentum transfer and is 
merely an approximation for collisions between the hard charged 
particles in a hot quark-gluon plasma. However, in our opinion it is a 
reasonable way to yield a first quantitative answer to the question of 
how collisions among hard particles affect the collective modes of 
QCD. The effects of such a collision term on the dispersion relations 
in the ultrarelativistic case for an isotropic system were 
investigated in \cite{Carrington:2003je}. For now the collision rate 
$\nu$ is taken to be a free parameter and we postpone the estimation 
of its magnitude to the discussions at the end of this paper.

Using (\ref{numbers}) we can write
\begin{equation}
V\cdot \partial_X \delta f^{i}_a(p,X) + g \theta_{i}
V_{\mu}F^{\mu\nu}_a(X)\partial_{\nu}^{(p)}f^{i}(\mathbf{p})=-\nu\left[f^{i}(\mathbf{p})+\delta
f^{i}_a(p,X)-\left(1+\frac{\int_{\mathbf{p}} \delta
f^{i}_a(p,X)}{N^{i}_{\text{eq}}}\right)
    f^{i}_{\text{eq}}(|\mathbf{p}|)\right]\text{\,.}\label{trans3}
\end{equation}
Solving for $\delta f^{i}_a(p,X)$ and Fourier-transforming leads to
the result for the linearized induced current by each particle species $i$ (see Appendix \ref{app1}):
\begin{align}
    J^{\mu\,i}_{\text{ind}\,a}(K)&=g^2 \int_{\mathbf{p}} V^{\mu}
    \partial^{\beta}_{(p)}
    f^{i}(\mathbf{p})\mathcal{M}_{\gamma\beta}(K,V)D^{-1}(K,\mathbf{v},\nu)A^{\gamma}_a(K) +g\nu \mathcal{S}^{i}(K,\nu) \notag\\
    &~~~~+g \frac{i \nu}{N^{i}_{\text{eq}}}\int_{\mathbf{p}}
    V^{\mu}f^{i}_{\text{eq}}(|\mathbf{p}|)D^{-1}(K,\mathbf{v},\nu)\notag\\
    &~~~~\times g \left[\int_{\mathbf{p}^{\prime}}\partial^{\beta}_{(p^{\prime})}f^{i}(\mathbf{p}^{\prime})
    \mathcal{M}_{\gamma\beta}(K,V^{\prime})D^{-1}(K,\mathbf{v}^{\prime},\nu)A^{\gamma}_a(K)+ g \nu \mathcal{S}^{i}(K,\nu)\right]\mathcal{W}_{i}^{-1}(K,\nu)\,\text{,}\label{current}
\end{align}
with 
\begin{equation}
\mathcal{M}_{\gamma\beta}(K,V):=g_{\gamma\beta}(\omega-\mathbf{k}\cdot\mathbf{v})-V_{\gamma}K_{\beta}\,\text{,}
\end{equation}
\begin{equation}
D(K,\mathbf{v},\nu):=\omega+i\nu-\mathbf{k}\cdot\mathbf{v}\,\text{,}
\end{equation}
\begin{equation}
\mathcal{S}^{i}(K,\nu):=\theta_{i} \int_{\mathbf{p}} V^{\mu}[f^{i}(\mathbf{p})-f^{i}_{\text{eq}}(|\mathbf{p}|)]D^{-1}(K,\mathbf{v},\nu)\,\text{,}
\end{equation}
and
\begin{equation}
\mathcal{W}_{i}(K,\nu):=1-\frac{i \nu}{N^{i}_{\text{eq}}}\int_{\mathbf{p}}f^{i}_{\text{eq}}(|\mathbf{p}|)D^{-1}(K,\mathbf{v},\nu)\text{\,.}
\end{equation}
The total induced current is given by
$J^{\mu}_{\text{ind}\,a}(K)=2 N_c J^{g\,\mu}_{\text{ind}\,a}(K) + N_f \left[J^{q\,\mu}_{\text{ind}\,a}(K)+
J^{\bar{q}\,\mu}_{\text{ind}\,a}(K)\right]$. It
can be simplified due to the fact that the integrals over
$f^{i}_{\text{eq}}$ in Eq.~(\ref{current}) are independent of the
particle species:
\begin{align}
    \frac{1}{N^{i}_{\text{eq}}}\int_{\mathbf{p}}f^{i}_{\text{eq}}(|\mathbf{p}|)D^{-1}(K,\mathbf{v},\nu)&=
    \int \frac{d\Omega}{4\pi}D^{-1}(K,\mathbf{v},\nu)\,\text{,}\notag\\
    \frac{1}{N^{i}_{\text{eq}}}\int_{\mathbf{p}}
    V^{\mu}f^{i}_{\text{eq}}(|\mathbf{p}|)D^{-1}(K,\mathbf{v},\nu)&=\int
    \frac{d\Omega}{4\pi}V^{\mu}D^{-1}(K,\mathbf{v},\nu)\,\text{,}\label{simpl}
\end{align}
where $d\Omega=\sin\theta d\theta d\varphi$. Assuming equal
distributions for quarks and antiquarks, the full linearized induced current
reads
\begin{align}
    J^{\mu}_{\text{ind}\,a}(K)&=g^2 \int_{\mathbf{p}}V^{\mu}
    \partial^{\beta}_{(p)}
    f(\mathbf{p})\mathcal{M}_{\gamma\beta}(K,V)D^{-1}(K,\mathbf{v},\nu)A^{\gamma}_a
    + 2 N_c g \nu \mathcal{S}^{g}(K,\nu)\notag\\
    &~~+ g^2 (i \nu) \int \frac{d\Omega}{4\pi}V^{\mu}D^{-1}(K,\mathbf{v},\nu)
    \int_{\mathbf{p}^{\prime}}\partial^{\beta}_{(p^{\prime})}f(\mathbf{p}^{\prime})
    \mathcal{M}_{\gamma\beta}(K,V^{\prime})D^{-1}(K,\mathbf{v}^{\prime},\nu)\mathcal{W}^{-1}(K,\nu)
    A^{\gamma}_a \notag\\
    &~~ + 2 N_c g^2 (i \nu^2)  \int \frac{d\Omega}{4\pi}V^{\mu}D^{-1}(K,\mathbf{v},\nu)
    \mathcal{S}^{g}(K,\nu) \mathcal{W}^{-1}(K,\nu)\,\text{,}\label{fullcurrent}
\end{align}
where $\mathcal{W}(K,\nu)=1-i \nu \int \frac{d\Omega}{4\pi}D^{-1}(K,\mathbf{v},\nu)$ 
and $f(\mathbf{p})$ as in Eq.~(\ref{f}). 
The self energy is obtained from Eq.~(\ref{fullcurrent}) via
\begin{equation}
    \Pi^{\mu\nu}_{ab}(K)=\frac{\delta J^{\mu}_{\text{ind}\,a}(K)}{\delta
    A_{\nu}^b(K)}\,\text{,}
\end{equation}
resulting in
\begin{align}
    \Pi^{\mu\nu}_{ab}(K)&=\delta_{ab} g^2 \int_{\mathbf{p}} V^{\mu}
    \partial_{\beta}^{(p)}
    f(\mathbf{p})\mathcal{M}^{\nu\beta}(K,V)D^{-1}(K,\mathbf{v},\nu)\notag\\
    &~~~+\delta_{ab} g^2 (i \nu)\int \frac{d\Omega}{4\pi}V^{\mu}D^{-1}(K,\mathbf{v},\nu)\int_{\mathbf{p}^{\prime}}
    \partial_{\beta}^{(p^{\prime})}f(\mathbf{p}^{\prime})
    \mathcal{M}^{\nu\beta}(K,V^{\prime})D^{-1}(K,\mathbf{v}^{\prime},\nu)\mathcal{W}^{-1}(K,\nu)
    \,\text{,}\label{selfenergy}
\end{align}
which is diagonal in color and can be shown to be transverse, i.e.,
$K_{\mu}\Pi^{\mu\nu}=K_{\nu}\Pi^{\mu\nu}=0$. We will from now on omit the color indices of $\Pi^{\mu\nu}$.
The terms in the induced current involving $\mathcal{S}_{\nu}$ drive the distribution into an isotropic
equilibrium shape. They represent a parity conserving current and thus create a
zero average electromagnetic field and therefore does not contribute to instability
growth.

\section{Dispersion relations}
\label{dispersion} In the linear approximation, the current that
is induced by the fluctuations can be expressed in terms of the
self energy by
\begin{equation}
    J^{\mu}_{\text{ind}\,1}(K)=\Pi^{\mu\nu}(K)A_{\nu}(K)\,\text{,}
\end{equation}
where
$J^{\mu}_{\text{ind}}=J^{\mu}_{\text{ind}\,1}+J^{\mu}_{\text{ind}\,2}$
and $J^{\mu}_{\text{ind}\,2}$ is the parity conserving part of the
current (\ref{fullcurrent}) that does not couple to the gauge
field.  Note that for a parity symmetric distribution as we use below,
$J^{\mu}_{\text{ind}\,2}=0$.
Inserting this into Maxwell's equation
\begin{equation}
    iK_{\mu}F^{\mu\nu}(K)=J^{\nu}_{\text{ind}\,1}+J^{\nu}_{\text{ext}}\,\text{,}
\end{equation}
we obtain
\begin{equation}
    \left[K^2g^{\mu\nu}-K^{\mu}K^{\nu}+\Pi^{\mu\nu}(K)\right]A_{\mu}(K)=J^{\nu}_{\text{ext}}(K)\,\text{,}
\end{equation}
with the external current $J^{\nu}_{\text{ext}}$. In the temporal
axial gauge, where $A_0=0$, this becomes
\begin{equation}
    \left[(k^2-\omega^2)\delta^{ij}-k^ik^j+\Pi^{ij}(K)\right]E^j(K)=[\Delta^{-1}(K)]^{ij}E^{j}(K)=i\omega
    J^{i}_{\text{ext}}(K)\,\text{,}\label{prop}
\end{equation}
and the response of the system to the external source is given by
\begin{equation}
    E^i(K)=i \omega \Delta^{ij}(K)J^{j}_{\text{ext}}(K)\text{\,.}
\end{equation}
The dispersion relations are obtained by finding the poles of the
propagator $\Delta^{ij}(K)$.

\section{Self energy and propagator in an anisotropic system}
\label{anisosystem} 

So far, the distribution function
$f(\mathbf{p})$ in Eq.~(\ref{selfenergy}) has not yet been
specified. We assume that $f(\mathbf{p})$ can be obtained from any
isotropic distribution function by rescaling one direction in
momentum space by defining \cite{Romatschke:2003ms}
\begin{equation}
    f(\mathbf{p})=\sqrt{1+\xi}\,f_{\text{iso}}\left(\mathbf{p}^2+\xi(\mathbf{p}\cdot\mathbf{\hat{n}})\right)\,\text{,}
    \label{anisodist}
\end{equation}
for an arbitrary isotropic distribution function
$f_{\text{iso}}(\mid\!\!\mathbf{p}\!\!\mid)$.  Distributions like (\ref{anisodist}) with
$\xi > 0$ are generated during a heavy-ion collision at times $\tau > \langle p_T \rangle^{-1}$.
The direction of the anisotropy above is
given by $\mathbf{\hat{n}}$ and $\xi>-1$ is an adjustable anisotropy
parameter. $\xi>0$ corresponds to a contraction of the
distribution in the $\mathbf{\hat{n}}$ direction, whereas $-1<\xi<0$
represents a stretching of the distribution in the $\mathbf{\hat{n}}$
direction. The factor of $\sqrt{1+\xi}$ ensures that the overall
particle number is the same for both the anisotropic and the
isotropic distribution function. With this particular distribution
we are able to perform the radial part of the integrations
involving $f(\mathbf{p})$ in Eq.~(\ref{selfenergy}) by changing
variables to
$\tilde{p}^2=p^2(1+\xi(\mathbf{v}\cdot\mathbf{\hat{n}})^2)$. The result
is
\begin{align}
    \Pi^{ij}(K)&=m_D^2\sqrt{1+\xi}\int\frac{d\Omega}{4\pi}v^i\frac{v^l+\xi(\mathbf{v}\cdot\mathbf{\hat{n}})n^l}{(1+\xi(\mathbf{v}\cdot\mathbf{\hat{n}})^2)^2}
    \left[\delta^{jl}(\omega-\mathbf{k}\cdot\mathbf{v})+v^jk^l\right]D^{-1}(K,\mathbf{v},\nu)\notag\\
    &~~~~+(i\nu) m_D^2 \sqrt{1+\xi} \int\frac{d\Omega^{\prime}}{4\pi}
    (v^{\prime})^i D^{-1}(K,\mathbf{v}^{\prime},\nu)\notag\\
    &~~~~\times\int\frac{d\Omega}{4\pi}\frac{v^l+\xi(\mathbf{v}\cdot\mathbf{\hat{n}})n^l}{(1+\xi(\mathbf{v}\cdot\mathbf{\hat{n}})^2)^2}
    \left[\delta^{jl}(\omega-\mathbf{k}\cdot\mathbf{v})+v^jk^l\right]D^{-1}(K,\mathbf{v},\nu)\mathcal{W}^{-1}(K,\nu)\,\text{,}
\end{align}
where
\begin{equation}
    m_D^2=-\frac{g^2}{2\pi^2}\int_0^{\infty}d\tilde{p}\,\tilde{p}^2\frac{d f_{\text{iso}}(\tilde{p}^2)}{d\tilde{p}}\text{\,.}
\end{equation}
We now decompose the self energy into four structure functions,
using the general tensor basis for an anisotropic system,
developed in \cite{Romatschke:2003ms}, such that
\begin{equation}
    \Pi^{ij}=\alpha A^{ij}+\beta B^{ij}+\gamma C^{ij}+\delta
    D^{ij}\,\text{,}
\end{equation}
where
\begin{equation}
    A^{ij}=\delta^{ij}-k^ik^j/k^2 \text{, } B^{ij}=k^ik^j/k^2
    \text{, } C^{ij}=\tilde{n}^i\tilde{n}^j/\tilde{n}^2 \text{, }
    D^{ij}=k^i\tilde{n}^j+k^j\tilde{n}^i \,\text{,}
\end{equation}
with $\tilde{n}^i=A^{ij}n^j$ the part of $n$ that is perpendicular
to $k$, i.e., $\tilde{n}\cdot k=0$. We determine the structure
functions by taking the contractions:
\begin{align}
    k^i\Pi^{ij}k^j=k^2\beta \text{, }
    \tilde{n}^i\Pi^{ij}k^j=\tilde{n}^2k^2\delta\text{, }
    \tilde{n}^i\Pi^{ij}\tilde{n}^j=\tilde{n}^2(\alpha+\gamma)\text{, }
    \text{Tr}\,\Pi^{ij}=2\alpha+\beta+\gamma\text{\,.}\label{contractions}
\end{align}
In this basis, the inverse of the propagator in Eq.~(\ref{prop})
can be written as
\begin{equation}
    \mathbf{\Delta}^{-1}(K)=(k^2-\omega^2+\alpha)\mathbf{A}+(\beta-\omega^2)\mathbf{B}+\gamma
    \mathbf{C}+\delta \mathbf{D}\,\text{,}
\end{equation}
while the propagator itself is given by \cite{Romatschke:2003ms}:
\begin{equation}
    \mathbf{\Delta}(K)=\Delta_A(\mathbf{A}-\mathbf{C})+\Delta_G\left[(k^2-\omega^2+\alpha+\gamma)\mathbf{B}+(\beta-\omega^2)\mathbf{C}-\delta\mathbf{D}\right]\,\text{,}
    \label{propagator}
\end{equation}
with
\begin{align}
    \Delta_A^{-1}(K)&=k^2-\omega^2+\alpha\,\text{,}\\
    \Delta_G^{-1}(K)&=(k^2-\omega^2+\alpha+\gamma)(\beta-\omega^2)-k^2\tilde{n}^2\delta^2\text{\,.}
\end{align}
Since the growth rate of the filamentation instability is the
largest when the wave vector is parallel to the direction of the
anisotropy, i.e., $\mathbf{k}\parallel\mathbf{\hat{n}}$ (see e.g. \cite{Romatschke:2003ms,Romatschke:2004jh,Arnold:2002zm}), we concentrate on this particular case. Then
\begin{equation}
    \mathbf{k}\cdot\mathbf{v}=k\,\mathbf{\hat{n}}\cdot\mathbf{v}=k
    \cos\theta\,\text{,}
\end{equation}
and $\gamma$ and $\tilde{n}^2=1-(\mathbf{k}\cdot{\mathbf{\hat{n}}})^2$
vanish identically. To determine the poles of the propagator
(\ref{propagator}), and hence the dispersion relations, we are now
left with two separate equations for the $\alpha-$ and $\beta-$mode, respectively:
\begin{align}
    k^2-\omega^2+\alpha &= 0\,\text{,}\notag\\
    \beta-\omega^2 &= 0\text{\,.}\label{equationstosolve}
\end{align}
From Eqs.~(\ref{contractions}) we find $\alpha$ and $\beta$, which
correspond to $\Pi_T$ and $(\omega^2/k^2)\Pi_L$, respectively. The
integrals can be solved analytically and the final results
simplify to:
        \begin{align}
            \alpha(\omega,k,\xi,\nu)&=\frac{m_D^2}{4}\frac{\sqrt{1+\xi}}{k (1+\xi z^2)^2}\left\{(k(z^2-1)-iz\nu)(1+\xi z^2)
            -(z^2-1)(k z(1+\xi)-i\nu)\ln\left[\frac{z+1}{z-1}\right]\notag\right.\\
            &~~~\left.-\frac{i}{\sqrt{\xi}}\left[z\nu(1+(3+z^2(1-\xi))\xi)+ik(1-\xi+z^2(1+\xi(6+z^2(\xi-1)+\xi)))\right] \arctan\sqrt{\xi}\right\}\,\text{,}
            \label{alphaeq}\\
            \beta(\omega,k,\xi,\nu)&=m_D^2\sqrt{1+\xi}k(kz-i\nu)^2\bigg\{-2\sqrt{\xi}(1+z^2\xi)+(1+\xi)\Big(2z\sqrt{\xi}\ln\left[\frac{z+1}{z-1}\right]
            +2(z^2\xi-1)\arctan\sqrt{\xi}\Big)\bigg\}\notag\\
            &~~~\times\left(2\sqrt{\xi}(1+z^2\xi)^2k^2\left(2k-i\nu\ln\left[\frac{z+1}{z-1}\right]\right)\right)^{-1}\,\text{,}
             \label{betaeq}
        \end{align}
where we abbreviate $z=(\omega+i\nu)/k$.

\section{Stable modes}
\label{stable}
\begin{figure}[t]
    \begin{center}
        \includegraphics[height=5.7cm]{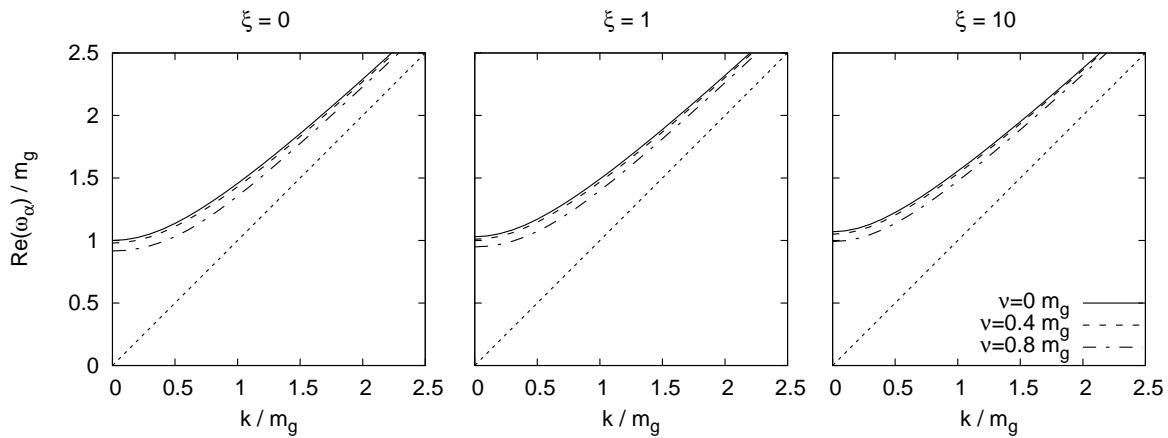}
        \caption{Real part of the dispersion relation for the stable $\alpha$-mode for an anisotropy parameter of $\xi=\{0,1,10\}$ and different collision rates in units of $m_g=m_D/\sqrt{3}$.}
        \label{fig:stablealphaxire}
    \end{center}
  \end{figure}

\begin{figure}[t]
    \begin{center}
        \includegraphics[height=4.4cm]{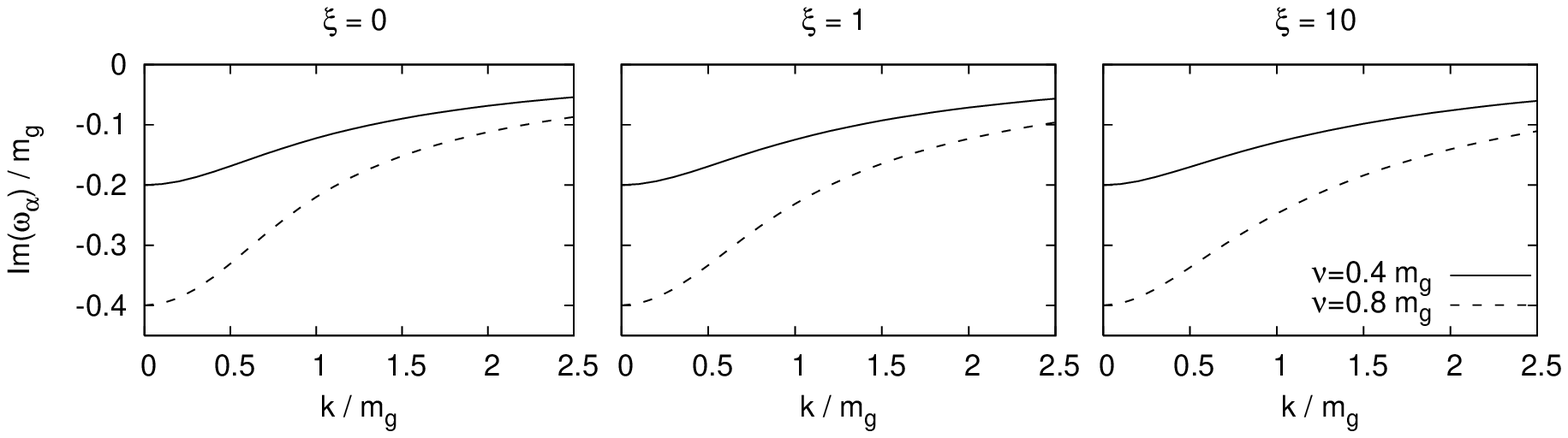}
        \caption{Imaginary part of the dispersion relation for the stable $\alpha$-mode for an anisotropy parameter of $\xi=\{0,1,10\}$ and different collision rates in units of $m_g=m_D/\sqrt{3}$.}
        \label{fig:stablealphaxiim}
    \end{center}
  \end{figure}

    The dispersion relations for all modes are given by the solutions $\omega(k)$ of Eqs.~(\ref{equationstosolve}).
    These solutions are found numerically for
    different values of the collision rate $\nu$.
    The results for the stable transverse ($\alpha$-) mode are
    shown in Figs.~\ref{fig:stablealphaxire} and
    \ref{fig:stablealphaxiim} for two
    anisotropic distributions, $\xi=1$ and
    $\xi=10$, together with the result for the isotropic case ($\xi=0$).
    Note that the finite collision rate
    causes $\omega_{\alpha}$ to become complex with a negative imaginary part corresponding to damping
    of these types of modes. 
    
\begin{figure}[t]
    \begin{center}
        \includegraphics[height=5.8cm]{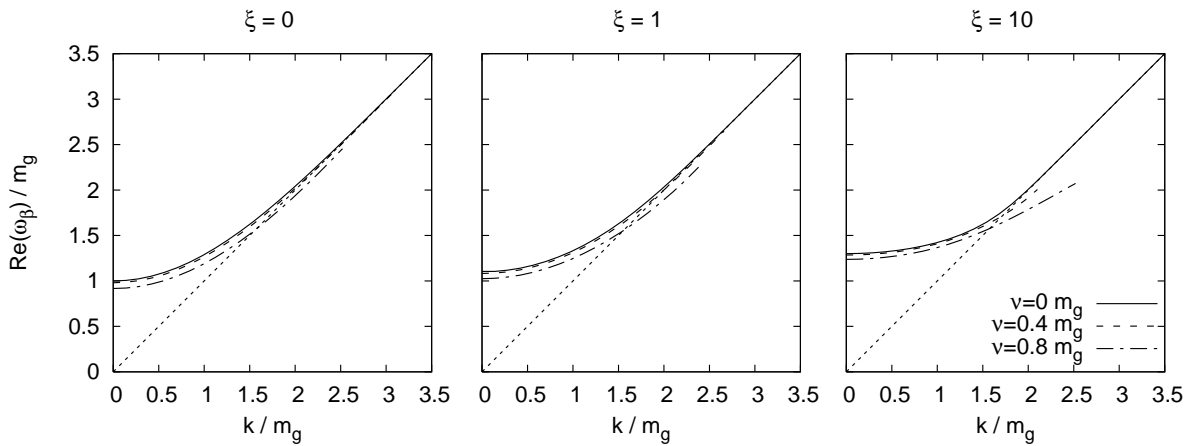}
        \caption{Real part of the dispersion relation for the stable $\beta$-mode for an anisotropy parameter of $\xi=\{0,1,10\}$ and
        different collision rates in units of $m_g=m_D/\sqrt{3}$.
        Note that for finite $\nu$ there is a maximal $k$ beyond which there is no solution.
        It vanishes when $\omega$ crosses the cut in the complex plane,
        which extends from $-k$ to $+k$ at $-i\nu$ (cf. Fig \ref{fig:stablebetaxiim}).}
        \label{fig:stablebetaxire}
    \end{center}
  \end{figure}

\begin{figure}[t]
    \begin{center}
        \includegraphics[height=4.6cm]{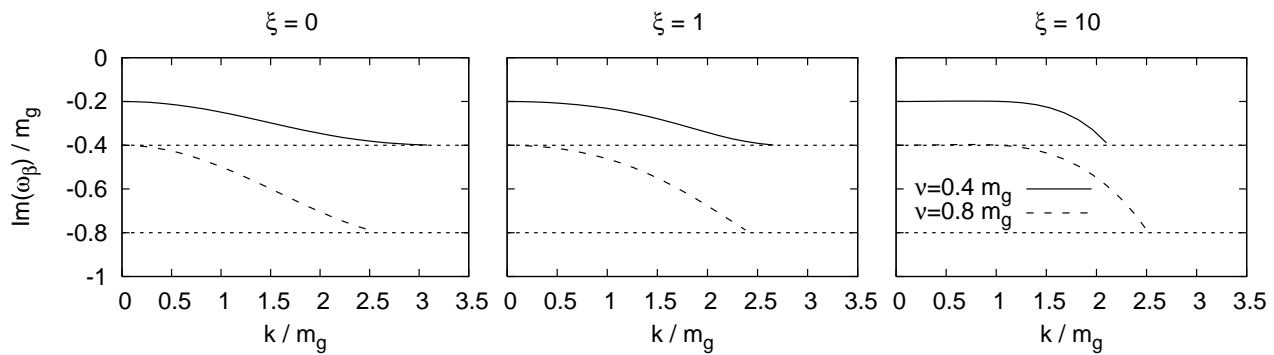}
        \caption{Imaginary part of the dispersion relation for the stable $\beta$-mode for an anisotropy parameter of $\xi=\{0,1,10\}$
        and different collision rates in units of $m_g=m_D/\sqrt{3}$.
        The solution vanishes, when $\omega$ crosses the cut in the complex plane, which extends from $-k$ to $+k$ at $-i\nu$ (indicated by the straight dotted line).}
        \label{fig:stablebetaxiim}
    \end{center}
  \end{figure}

  The effect of collisions on the stable longitudinal ($\beta$-) modes is more significant.
  The results are presented in Figs.~\ref{fig:stablebetaxire} and
  \ref{fig:stablebetaxiim}. We find that for finite $\nu$ the
  dispersion becomes spacelike (${\rm Re}(\omega)<k$) at large $k$ in contrast to
  the collisionless case, in which ${\rm Re}(\omega)>k$ always holds. This
  behavior is responsible for the fact that the solution vanishes from the physical Riemann sheet
  above some finite $k$. This occurs precisely when the solution for $\omega_{\beta}$ reaches the
  cut between $-k$ and $k$ at $-i\nu$.


\begin{figure}[t]
      \begin{center}
        \includegraphics[height=6cm]{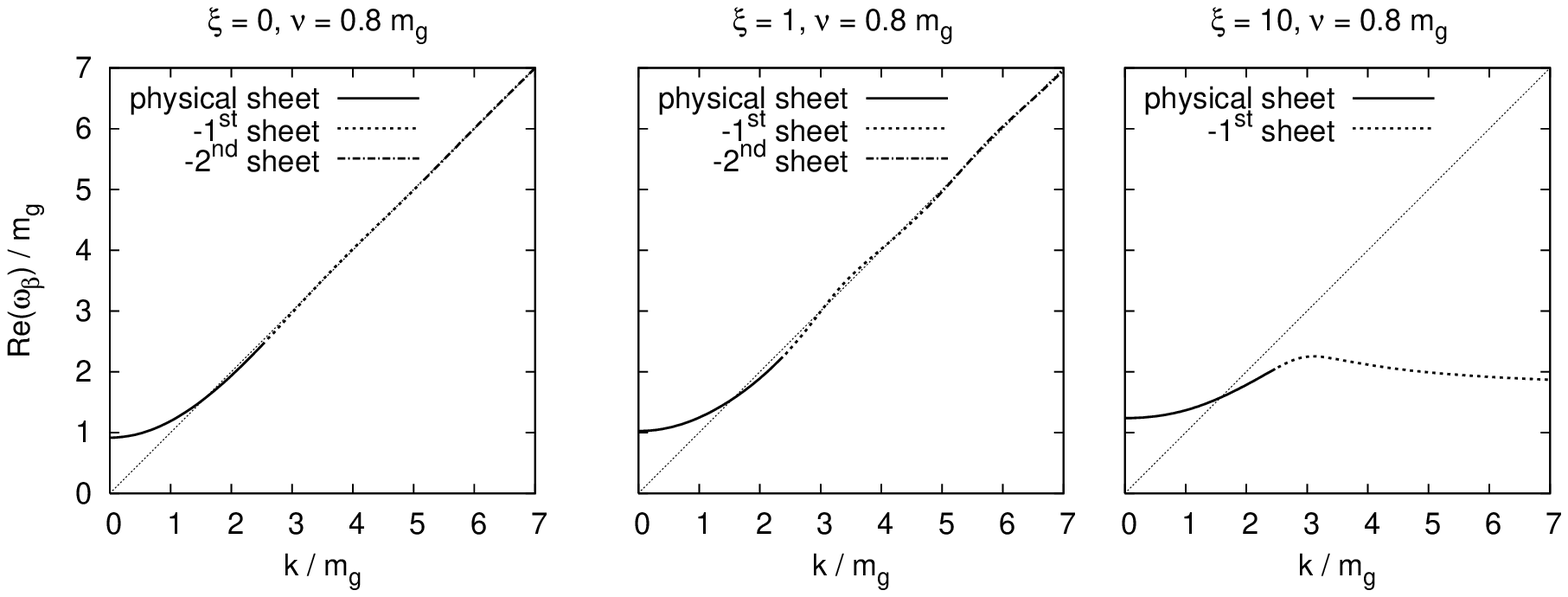}
        \caption{Real part of the dispersion relation for the stable $\beta$-mode for an anisotropy parameter of $\xi=\{0,1,10\}$ and
        $\nu=0.8\,m_g$. It is shown how the solution continues on lower Riemann
        sheets.}
        \label{fig:stablebetahigherxire}
    \end{center}
  \end{figure}

\begin{figure}[t]
      \begin{center}
        \includegraphics[height=4.7cm]{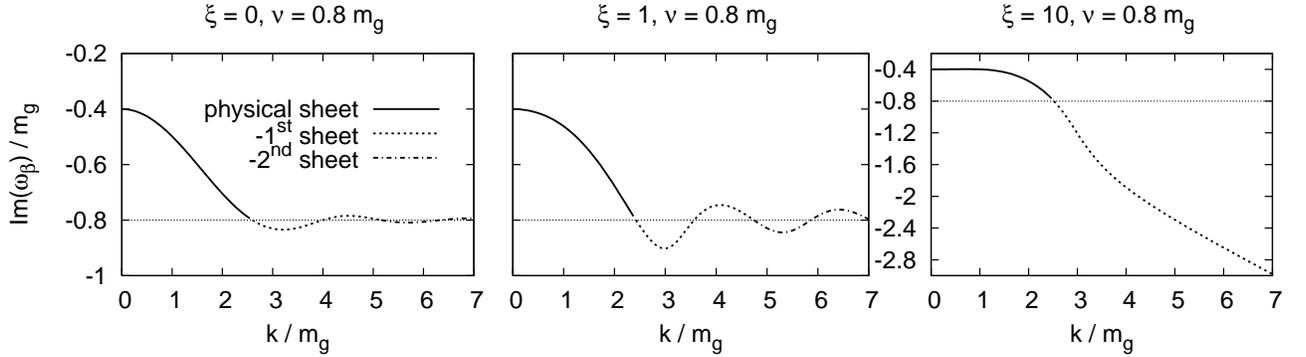}
        \caption{Imaginary part of the dispersion relation for the stable $\beta$-mode for an anisotropy parameter of $\xi=\{0,1,10\}$ and
        $\nu=0.8\,m_g$. It is shown how the solution continues on lower Riemann sheets. Note the different scale for the third plot.}
        \label{fig:stablebetahigherxiim}
    \end{center}
  \end{figure}

  It is however possible to
  find solutions on the unphysical Riemann sheets by replacing
  the logarithm in the structure function with its usual
  analytic continuation \cite{Romatschke:2004jh}:
  \begin{align}
    \ln\left(\frac{z+1}{z-1}\right)=\ln\left(\left|\frac{z+1}{z-1}\right|\right)+i\left[\arg\left(\frac{z+1}{z-1}\right)+2\pi
    N\right]\,\text{,}
  \end{align}
  where $N$ specifies the sheet number. The continuation of the
  solution to the lower Riemann sheets is shown in
  Figs.~\ref{fig:stablebetahigherxire} and
  \ref{fig:stablebetahigherxiim} for a collision rate of
  $\nu=0.8\,m_g \approx 0.46\,m_D$ and different anisotropy
  parameters $\xi$. For smaller anisotropies the solution is found
  to converge to the light cone in an oscillating manner, while
  the imaginary part of $\omega$ oscillates around $\pm\nu$. Between $\xi=2$ and
  $\xi=3$ (for $\nu \approx 0.46\,m_D$) this behavior changes qualitatively to the one shown in
  the case $\xi=10$. With increasing $k$ the real part of $\omega$ moves
  away from the light cone, while the imaginary part drops to large negative
  values with the solution remaining on the $N=-1$ Riemann sheet, such that these modes are strongly damped.

\section{Unstable modes}
\label{unstable}
    Anisotropic momentum distributions cause kinetic
    instabilities. For the particular distribution
    (\ref{anisodist}) with $\mathbf{\hat{n}}\parallel\mathbf{k}$ and $\xi>0$ there
    exists a magnetic instability, the so called filamentation or
    Weibel instability \cite{Weibel:1959}. Its existence is due to
    a surplus of particles with momentum perpendicular
    (or close to perpendicular) to $\mathbf{k}$.
    These particles are trapped in the direction of $\mathbf{k}$ by the background magnetic field
    and cause currents, which generate magnetic fields that add to the original one. Hence
    they contribute to instability,
    while all other particles have a stabilizing effect.
    (for a more detailed qualitative review of this scenario see \cite{Arnold:2003rq}).
    In the isotropic case the stabilizing and destabilizing
    contributions cancel, such that no instability arises.

\begin{figure}[t]
  \hfill
  \begin{minipage}[t]{.45\textwidth}
      \begin{center}
        \includegraphics[height=5cm]{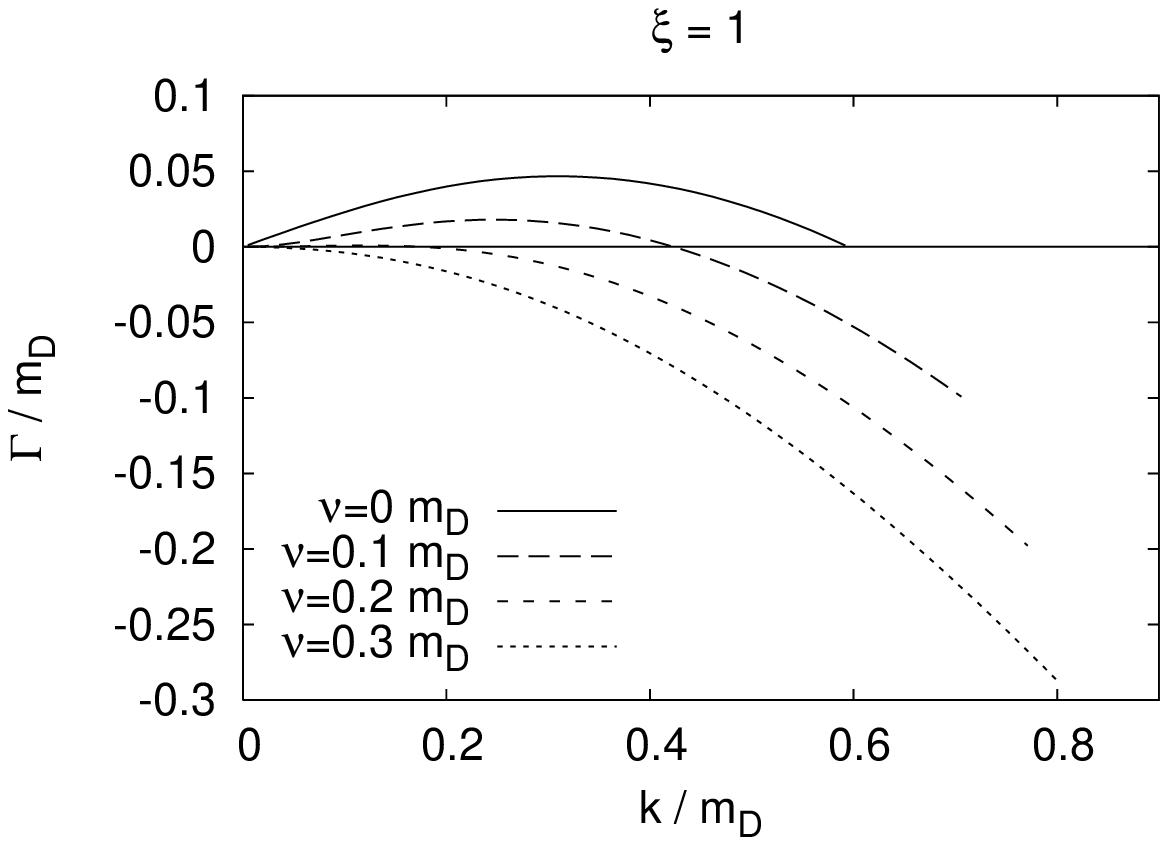}
        \caption{Dependence of the growth rate $\Gamma$ of the unstable transverse ($\alpha-$) mode on the collision rate $\nu$, for an anisotropy parameter $\xi=1$.}
        \label{fig:instablexi1}
      \end{center}
  \end{minipage}
  \hfill
  \begin{minipage}[t]{.45\textwidth}
      \begin{center}
        \includegraphics[height=5cm]{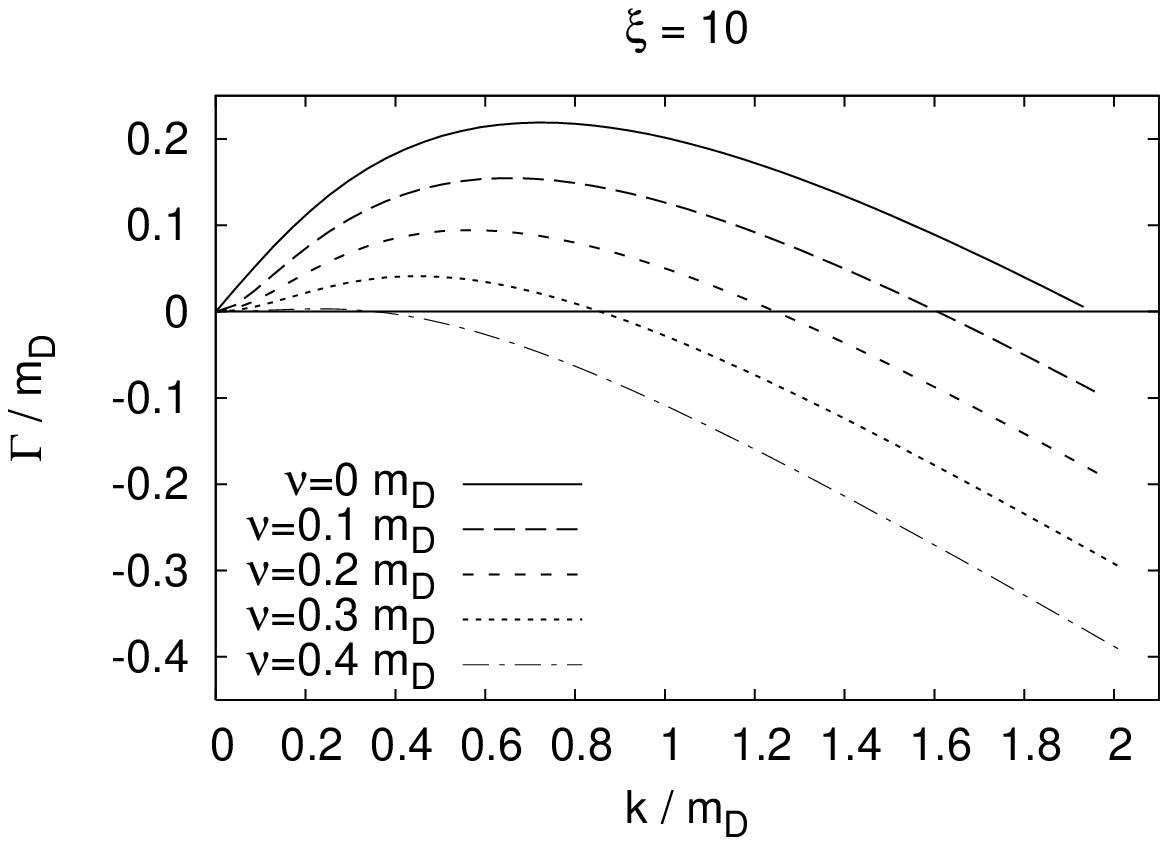}
        \caption{Dependence of the growth rate $\Gamma$ of the unstable transverse ($\alpha-$) mode on the collision rate $\nu$, for an anisotropy parameter $\xi=10$.}
        \label{fig:instablexi10}
      \end{center}
  \end{minipage}
  \hfill
\end{figure}

\begin{figure}[t]
  \hfill
  \begin{minipage}[t]{.45\textwidth}
      \begin{center}
        \includegraphics[height=5cm]{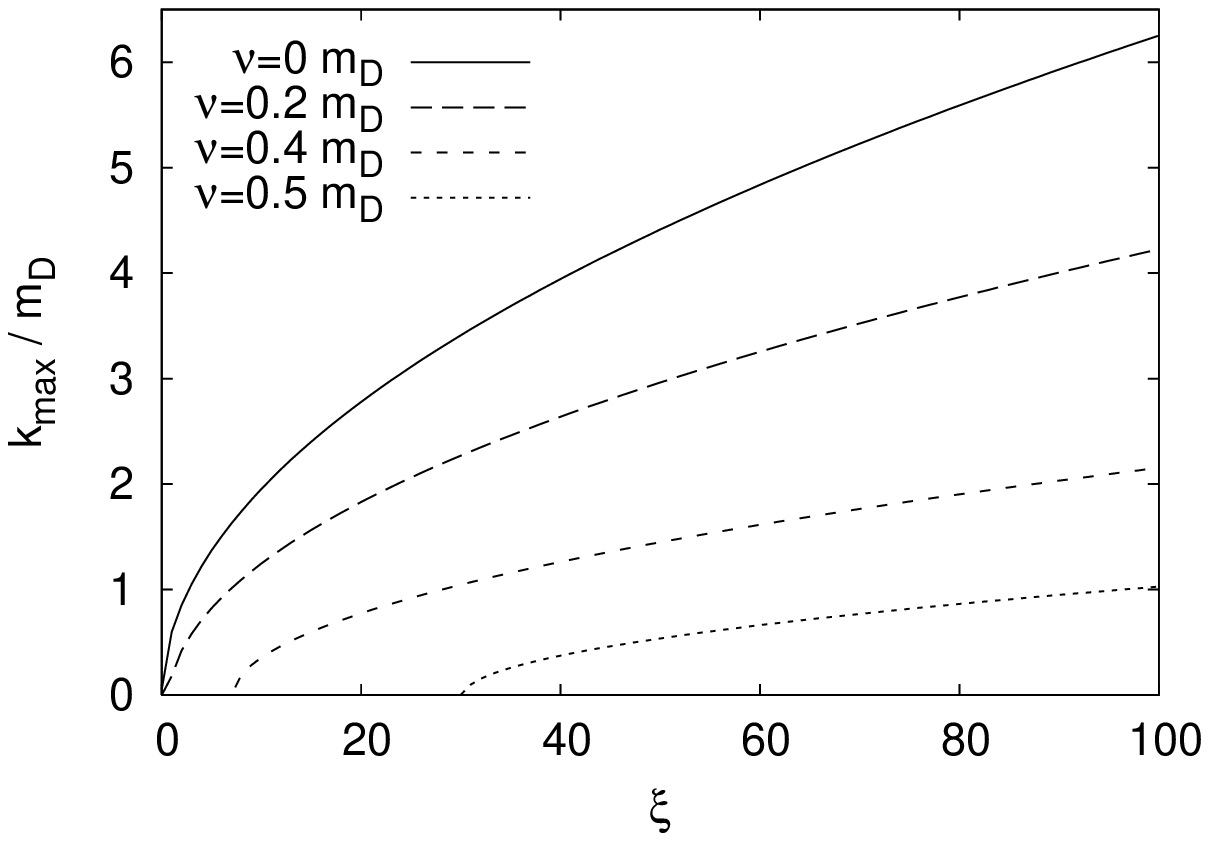}
        \caption{$k_{\text{max}}$ of the maximally unstable mode as a function of the anisotropy parameter $\xi$ for different values of $\nu$.
        For given $\nu$, the momentum anisotropy parameter must be above the critical value for instabilities to occur.}
        \label{fig:kmaxofxi}
      \end{center}
  \end{minipage}
  \hfill
  \begin{minipage}[t]{.45\textwidth}
      \begin{center}
        \includegraphics[height=5cm]{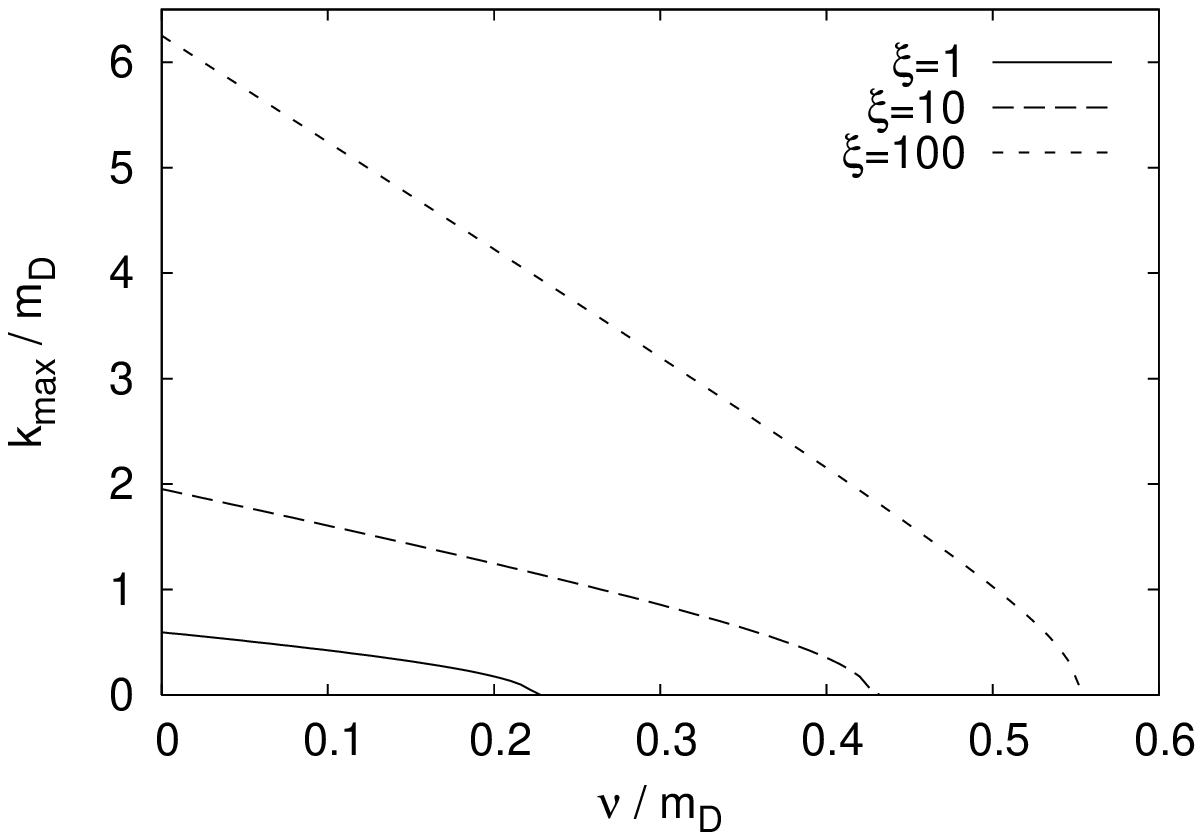}
        \caption{$k_{\text{max}}$ of the maximally unstable mode as a function of the collision rate $\nu$ for given anisotropy parameter $\xi$.
        There exist $\xi$-dependent critical collision rates $\nu_{\text{max}}(\xi)$ above which instabilities can not exist.}
        \label{fig:kmaxofnu}
      \end{center}
  \end{minipage}
  \hfill
\end{figure}

    We now investigate how the inclusion of collisions as
    described in Section \ref{includecollisions} affects the
    growth rates of these instabilities. Qualitatively one
    expects a decrease of the growth rates because the particles,
    which move perpendicular to $\mathbf{k}$, can scatter with
    other particles and will no longer be trapped. Other particles
    can gain a momentum close to perpendicular to
    $\mathbf{k}$ and form a new contribution to the instability.
    However, since the collision term tends to randomize the
    momentum distribution, the growth of $\delta f$ and the magnetic field
    is prevented.
    In order to describe this effect quantitatively, we
    solve Eq.~(\ref{equationstosolve}) for purely imaginary
    $\omega$ and vary the collision rate $\nu$. The solution
    $\omega(k)=i\Gamma(k)$ gives the growth rate $\Gamma(k)$.
    In the case that $\mathbf{\hat{n}}\parallel\mathbf{k}$
    solutions like that only exist for the transverse ($\alpha$-) mode.
    The one with $\Gamma>0$ corresponds to the filamentation instability.
    Results for different values of the collision rate $\nu$ are
    shown in Fig.~\ref{fig:instablexi1} for $\xi=1$ and in
    Fig.~\ref{fig:instablexi10} for $\xi=10$.
    The qualitatively expected effect is nicely reproduced. The
    growth rate decreases with an increasing collision rate as
    does the maximal wave number for an unstable mode.
    One can see that in
    the case where $\xi=1$ already for $\nu$ being around $20\%$ of the Debye mass,
    growth has completely turned into damping and no instability
    can evolve. For $\xi=10$ the collision rate $\nu$ has to be
    slightly larger than $30\%$ of the Debye mass in order to
    prevent growth of a collective mode.
In order to find the wave number $k_{\text{max}}(\xi,\nu)$ at which
the unstable mode spectrum terminates, we take 
the limit $\omega \rightarrow 0$ to obtain
\begin{align}
    m_{\alpha}^2=\lim_{\omega \rightarrow 0} \alpha &=
    -\frac{m_D^2}{8} i k
    \frac{\sqrt{1+\xi}}{\sqrt{\xi}(k^2-\nu^2\xi)^2}\left.\bigg\{-2 i k
    \sqrt{\xi}(k^2-\nu^2\xi)\right.\notag\\
    & ~~~~~~~\left.-2\nu
    (k^2+\nu^2)\xi^{3/2}\ln\left(1-\frac{2k}{k-i\nu}\right)
    -2 i
    k\left[k^2(\xi-1)+\nu^2\xi(\xi+3)\right]\arctan(\sqrt{\xi})\right.\bigg\}\text{\,.}\label{ma2}
\end{align}
One of the solutions to the equation
\begin{equation}
    k^2+m_{\alpha}^2=0\,\text{,}\label{kmaxeq}
\end{equation}
which is just the limit $\omega\rightarrow 0$ of the first of the
Eqs.~(\ref{equationstosolve}), is $k_{\text{max}}$. Results for
different $\xi$ and $\nu$ are shown in Figs.~\ref{fig:kmaxofxi}
and \ref{fig:kmaxofnu}. We find that for a given anisotropy
parameter $\xi$, there exists a critical collision rate, above which
instabilities can not occur. This is also true for the limit $\xi
\rightarrow \infty$, as we will show in the following. Taking 
$\xi\rightarrow \infty$, Eq.~(\ref{ma2}) becomes
\begin{equation}
    m_{\alpha}^2(\xi\rightarrow
    \infty)=-\frac{\pi}{8}m_D^2\frac{k^2}{\nu^2}\,\text{,}
\end{equation}
which together with Eq.~(\ref{kmaxeq}) gives
\begin{align}
    k^2(\nu^2-\frac{\pi}{8}m_D^2)&=0\text{\,.}
\end{align}
Apart from the result $k=0$, this is solved by
\begin{equation}
    \nu_{\text{max}}(\xi\rightarrow\infty)=\sqrt{\frac{\pi}{8}}m_D\approx 0.6267\, m_D =\Gamma_{\text{max}}(\xi\rightarrow\infty)\,\text{,}
    \label{numaxeq}
\end{equation}
the critical collision rate, above which even in the extremely
anisotropic limit $\xi=\infty$ no instability can occur. 

\begin{figure}[t]
      \begin{center}
        \includegraphics[height=5cm]{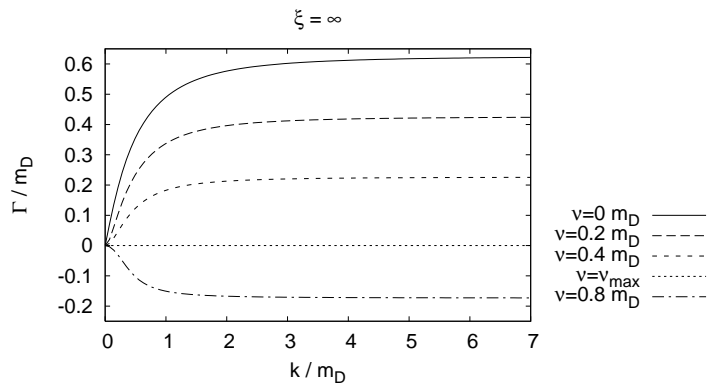}
        \caption{Dependence of the growth rate $\Gamma$ of the unstable transverse ($\alpha-$) mode on the collision rate $\nu$, for the extremely anisotropic limit
        $\xi=\infty$.}
        \label{fig:unstablealphaxiinf}
      \end{center}
  \end{figure}

\begin{figure}[t]
      \begin{center}
        \includegraphics[height=6cm]{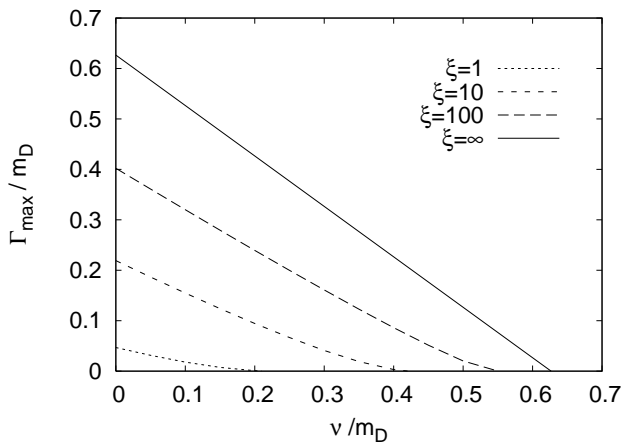}
        \caption{The maximal growth rate of the instability as a function of the collision rate $\nu$.}
        \label{fig:gammamax}
      \end{center}
  \end{figure}

This is
also visible in the plot of the growth rate
$\Gamma_{\xi\rightarrow\infty}$, shown in
Fig.~\ref{fig:unstablealphaxiinf}. For
$\nu=\nu_{\text{max}}(\xi\rightarrow\infty)$ the growth rate
becomes zero for all $k$, and for larger $\nu$ only damping
occurs. In this case ($\xi=\infty$) the value of the maximal collision 
rate equals that of the maximal growth rate in the collisionless limit. 
This simply means that the instability vanishes completely at the point where
the collisions damp at the same rate at which the instability grows. Note however
that this relation is more complicated in general as shown in Fig.~\ref{fig:gammamax}, 
where the dependence of the maximal growth rate on the collision rate is plotted. 
In order to make the instability vanish
completely for any $\xi<\infty$, a collision rate larger than the maximal growth rate of the instability in 
the collisionless limit is needed.


\section{Discussions}
\label{discussions}

In the body of the text the collisional frequency $\nu$ has been taken 
to be arbitrary. Because the inclusion of a BGK collisional term is a 
phenomenological model for the equilibration of a system and cannot be 
derived from first principles this makes it difficult to fix the 
magnitude of $\nu$.  However, since as treated here the underlying 
framework of Boltzmann-Vlasov is implicitly perturbative, we can 
attempt to fix $\nu$ perturbatively.  However, even this is non-trivial 
since within non-abelian theories there are at least two possible 
collisional frequencies to be considered 
\cite{Selikhov:1993ns,Bodeker:1998hm,Arnold:1998cy,Bodeker:1999ey,Arnold:2002zm}: 
(1) the frequency for hard-hard 
scatterings which is parametrically $\nu_{\rm hard} \sim 
\alpha_s^2 \log \alpha_s^{-1}$ and (2) the frequency for hard-soft 
scattering which is parametrically $\nu_{\rm soft} \sim \alpha_s \log 
\alpha_s^{-1}$.

The hard-hard scatterings correspond to interactions which change the 
momentum of a hard particle by ${\cal O}(p_{\rm hard})$ and therefore 
represent truly momentum-space isotropizing interactions.  On the 
other hand the hard-soft scatterings correspond to changes in momentum 
which are order ${\cal O}(g p_{\rm hard})$. These small angle 
scatterings occur rather frequently and it turns out that after 
traversing one hard scattering mean free path, $\lambda_{\rm hard} 
\sim \nu_{\rm hard}^{-1}$, the typical deflection of the particle is 
also ${\cal O}(1)$.\footnote{This, in the end, is the source of the 
logarithm in $\nu_{\rm hard}$ above.}  The physics of small-angle 
scattering by the soft-background is precisely what is captured by the 
hard-loop treatment.  However, the hard-loop framework doesn't 
explicitly take into account that $\nu_{\rm soft}$ is also the 
frequency at which there are color-rotating interactions of the hard 
particles themselves. One would expect that color-rotation of the hard 
particles to have a larger effect on the growth of instabilities than 
the momentum-space isotropization via hard-hard scattering.  That 
being said, the form of the BGK scattering kernel does not mix color 
channels and in that sense cannot be used to describe the physics of 
color-rotation of the hard particles.  For this reason one is lead to 
the conclusion that when using the BGK kernel the appropriate 
damping rate is $\nu \sim \nu_{\rm hard} \sim \alpha_s^2 \log \alpha_s^{-
1}$.  This conclusion for the parametric dependence of $\nu$ is also
supported by looking at the leading order result for the shear
viscosity \cite{Thoma:1993vs}.

Even with this conclusion it is hard to say anything quantitative 
about $\nu$ since the overall coefficient and the coefficient in the 
logarithm are not specified by such a parametric relation.  One could 
hope that previous calculations contained in \cite{Thoma:1993vs} of 
parton interaction rates could be of some use.  Unfortunately, for a 
purely gluonic plasma it was found $\nu = 5.2 \, \alpha_s^2 T \, 
\log(0.25 \alpha_s^{-1})$, which clearly cannot be trusted for the 
values of $\alpha_s$ which are relevant for heavy-ion collisions 
($\alpha_s \sim 0.2-0.4$) since $\nu$ becomes negative for large 
$\alpha_s$. We note here that the negativity of this result at 
large couplings most likely stems from the strict perturbative 
expansion of the integrals involved failing when the hard and soft 
scales become comparable in magnitude. Similar erroneous negative 
values also occur in the perturbative expressions for heavy-quark 
collisional energy loss \cite{Braaten:1991we} when extrapolated to large coupling. 
A corrected calculational method which yields positive-definite 
results for the heavy-quark energy loss was detailed in 
Refs.~\cite{Romatschke:2003vc, Romatschke:2004au}. 

Ideally, one would revisit the calculation of the interaction rate 
and improve upon the techniques used where necessary.  Short of such a 
calculation one cannot say with certainty what the numerical value of 
$\nu$ should be and the best we can do is to play ``games''.  For 
instance, one could insert a one into the logarithm appearing in $\nu$ similar to what other
authors have done \cite{Kapusta:1991qp} to obtain 
$\nu \sim 5.2 \, \alpha_s^2 T \log(1 + 0.25 \alpha_s^{-1}) \sim 0.1-0.2\,m_D$ for $\alpha_s = 
0.2-0.4$.  Using this admittedly specious expression for large-coupling one can 
expand and obtain $\lim_{\alpha_s \rightarrow \infty} 
\nu/m_D \sim 0.37\, \alpha_s^{1/2}$.  Note that in the case $N_f=2$ both $\nu$ and $m_D$ 
increase; however, the ratio of these two scales is still in the range 
quoted for $N_f=0$.  The range, $\nu = 0.1-0.2\,m_D$, places us well 
below the threshold needed to turn off instabilities in the case of 
extremely anisotropic distribution functions but does imply (see 
Fig.~\ref{fig:gammamax}) that for moderately small anisotropies, $\xi 
\sim 1$, and large coupling that it is possible for collisional 
damping to eliminate the unstable modes from the spectrum completely. 
Of course, in the limit of asymptotically small couplings the ratio 
$\nu/m_D$ approaches zero and the collisionless results hold to very 
good approximation.  In the opposite limit of strong coupling the estimates
here are at best guesswork and it is indeed possible that the ratio
$\nu/m_D$ is larger than the range we have quoted.  For example, the
recent work of Peshier \cite{Peshier:2004bv,Peshier:2005pp} implies that
$\nu/m_D$ could be as large as 0.5; however, this number results from
a fit of a model assumption to lattice data and is not directly comparable
to the collisional widths considered here since in their description 
the gluon width was assumed to be parametrically given by $\nu_{soft}$.

Additionally, we have to mention the caveat that all the estimates
above rely on full equilibrium thermal field theory calculations.  For the
very initial state of the matter created in an ultrarelativistic heavy-ion collision
the system is clearly not in equilibrium and it is not clear how this estimate
will change as a result.  However, it is of crucial importance to attempt
to estimate the scattering rate in a non-equilibrium setting.

\section{Conclusions}
\label{conclusions}

In this paper we have studied the effects of including a BGK collision 
kernel on the collective modes of a QCD plasma which has a hard particle
distribution function which is anisotropic in momentum space.  To 
simplify the analysis we have specialized to gluonic collective modes 
which have their momentum vector, ${\bf k}$, directed along the 
anisotropy direction, $\hat{\bf n}$.  The reasons for doing this were 
two-fold:  (1) in the collisionless limit these modes correspond to 
the modes which are the most unstable and (2) when $\hat{\bf k} 
\parallel \hat{\bf n}$ it is possible to obtain analytic expressions, 
Eqs.~(\ref{alphaeq}) and (\ref{betaeq}), for the soft-gluon self-energy 
structure functions.  We have presented herein dispersion 
relations for both the stable and unstable modes in the case that 
there is a finite collisional frequency (or damping coefficient), 
$\nu$.  Our results confirm what can be expected intuitively, namely 
that the addition of collisional damping slows down the rate of growth 
of the unstable modes.  

However, going beyond this intuitive expectation we have presented 
detailed calculations of the dependence of the maximal unstable mode 
growth rate on the parameter $\nu$ as shown in 
Fig.~\ref{fig:gammamax}.  For all values of the anisotropy parameter, 
$\xi$, we find that there is a critical value of $\nu_{max}$ above 
which no instability is present in the system.  In Fig.~\ref{fig:gammamax} this 
corresponds to the value of $\nu$ at which the maximal growth rate 
vanishes.  In the limit that $\xi \rightarrow \infty$ we were able to 
derive an analytic expression for $\nu_{\max}$, Eq.~(\ref{numaxeq}), finding 
that it corresponds precisely to the maximal growth rate obtained in 
the collisionless limit.

In addition, we have investigated the non-trivial analytic structure 
of the soft-gluon propagator in this model finding that the stable 
longitudinal mode becomes spacelike\footnote{In the sense that ${\rm 
Re}(\omega_L/k) < 1$.} and only remains on the physical Riemann sheet 
up to a certain critical momentum.  Beyond this critical momentum the 
solution goes through the logarithmic cut to the $N=-1$ Riemann sheet. For 
weak-damping the longitudinal mode then continues to spiral around the 
logarithmic branch point onto lower and lower Riemann sheets as its 
momentum increases (see Figs.~\ref{fig:stablebetahigherxire}a and 
\ref{fig:stablebetahigherxiim}a and Figs 
\ref{fig:stablebetahigherxire}b and \ref{fig:stablebetahigherxiim}b). 
In the case of stronger anisotropies the longitudinal mode solution no 
longer ``spirals down'' the logarithmic branch point to lower Riemann 
sheets but instead simply moves lower in the complex plane of the $N=-
1$ Riemann sheet (see Figs.~\ref{fig:stablebetahigherxire}c and 
\ref{fig:stablebetahigherxiim}c).  We note that even in the isotropic 
case, the detailed analytic structure of the collisionally damped 
modes enables one to calculate quantities such as the QCD pressure 
including the effects of damping of the quasiparticle modes which could
be of some interest \cite{Andersen:1999fw,Andersen:1999sf,Andersen:2002ey,Andersen:2003zk,Peshier:2004bv,Peshier:2005pp}.

In the discussions section of the manuscript we have attempted to fix
a numerical value for the collisional frequency $\nu$.  We have argued
that since the BGK kernel does not rotate the color of the hard particles
that the appropriate frequency is parametrically given by the time scale
for hard-hard collisions, namely $\nu \sim \alpha_s^2 \log \alpha_s^{-1}$.
Going further than this parametric estimate to a value applicable at
couplings expected to be generated during heavy-ion collisions 
($\alpha_s \sim 0.2-0.4$) is problematic due to the small coefficient which
appears in the logarithm resulting in $\nu$ becoming negative.  Playing a game
by adding a one in the argument of the logarithm we found that 
$\nu \sim 0.1-0.2\,m_D$ which according
to the results of this paper would imply that for weak anisotropies there
are no instabilities (see Fig.~\ref{fig:gammamax}).  For stronger anisotropies
the results contained here tell us how much the maximal growth rate of the
unstable modes is affected by the inclusion of collisional damping via a BGK
kernel.  Looking forward more detailed calculations of collision rates in
a time-evolving soft-field background and true non-equilibrium situation
are clearly needed.

\section*{Acknowledgements}
We would like to thank Peter Arnold and Stanislaw Mrowczynski 
for useful discussions.

\appendix
\section{Analytical solution of the linearized transport equations}
\label{app1}
We briefly sketch how the result Eq.~(\ref{current}) emerges from the transport equation (\ref{trans3}).
From Eq.~(\ref{trans3}) we immediately get
\begin{align}
	(-i\omega+i\mathbf{v}\cdot\mathbf{k}+\nu)\delta f^{i}(p,K) = -g\theta_iV_{\mu}F^{\mu\nu}(K)\partial_{\nu}^{(p)}f^{i}(\mathbf{p})+\nu(f^{i}_{\text{eq}}(\mathbf{p})-f^{i}(\mathbf{p}))+\nu \frac{f^{i}_{\text{eq}}(\mathbf{p})}{N_{\text{eq}}} \int_{\mathbf{p}^{\prime}} \delta f^{i}(p^{\prime},K)\,\text{,}
\end{align}
where $\delta f^{i}(p,K)$ and $F^{\mu\nu}(K)$ are the Fourier-transforms of $\delta f^{i}(p,X)$ and $F^{\mu\nu}(X)$, respectively.
This yields
\begin{equation}
	\delta f^{i}(p,K)=\frac{-ig\theta_iV_{\mu}F^{\mu\nu}(K)\partial_{\nu}^{(p)}f^{i}(\mathbf{p})+i\nu(f^{i}_{\text{eq}}(\mathbf{p})-f^{i}(\mathbf{p}))+i\nu f^{i}_{\text{eq}}(\mathbf{p})\left( \int_{\mathbf{p}^{\prime}}\delta f^{i}(p^{\prime},K) \right)/N_{\text{eq}}}{\omega-\mathbf{v}\cdot\mathbf{k}+i\nu}\text{\,.}
\end{equation}
Defining
\begin{equation}
	\delta f^{i}_0(p,K)=\left(-ig\theta_iV_{\mu}F^{\mu\nu}(K)\partial_{\nu}^{(p)}f^{i}(\mathbf{p})+i\nu(f^{i}_{\text{eq}}(\mathbf{p})-f^{i}(\mathbf{p}))\right)D^{-1}(K,\mathbf{v},\nu)\,\text{,}
\end{equation}
with $D(K,\mathbf{v},\nu)=\omega-\mathbf{k}\cdot\mathbf{v}+i\nu$ we can write
\begin{align}
	\delta f^{i}(p,K)=\delta f^{i}_0(p,K)&+i\nu D^{-1}(K,\mathbf{v},\nu)\frac{f^{i}_{\text{eq}}(\mathbf{p})}{N_{\text{eq}}}\int_{\mathbf{p}^{\prime}}\delta f^{i}_0(p^{\prime},K)\notag\\
	&+i\nu D^{-1}(K,\mathbf{v},\nu)\frac{f^{i}_{\text{eq}}(\mathbf{p})}{N_{\text{eq}}} \frac{i\nu}{N_{\text{eq}}} \int_{\mathbf{p}^{\prime}} f^{i}_{\text{eq}}(\mathbf{p}^{\prime})
	D^{-1}(K,\mathbf{v}^{\prime},\nu)
	\int_{\mathbf{p}^{\prime\prime}}\delta f^{i}_0(p^{\prime\prime},K)\notag\\
	&+\ldots
\end{align}
Using the shorthand notation
\begin{equation}
\eta(K)=\int_{\mathbf{p}}\delta f^{i}_0(p,K)
\end{equation}
and
\begin{equation}
\lambda(K,\nu)=\frac{i\nu}{N_{\text{eq}}}\int_{\mathbf{p}}
	f^{i}_{\text{eq}}(\mathbf{p})D^{-1}(K,\mathbf{v},\nu)
\end{equation}
we finally have
\begin{align}
	\delta f^{i}(p,K)&=\delta f^{i}_0(p,K)+i\nu D^{-1}(K,\mathbf{v},\nu)\frac{f^{i}_{\text{eq}}(\mathbf{p})}{N_{\text{eq}}} \eta(K) \left(1+\lambda+\lambda^2+\ldots\right)\notag\\
&=\delta f^{i}_0(p,K)+i\nu D^{-1}(K,\mathbf{v},\nu)\frac{f^{i}_{\text{eq}}(\mathbf{p})}{N_{\text{eq}}} \eta(K)\frac{1}{1-\lambda}\,\text{,}
\end{align}
which translates to the final result for the current (\ref{current}) by using
\begin{equation}
	J_{\text{ind}\,a}^{\mu\, i}(K)=g \int_{\mathbf{p}} V^{\mu} \delta f^{i}_a(p,K)\,\text{,}
\end{equation}
where we reintroduced the color indices.

\bibliography{anisocollisions}

\end{document}